\documentclass[prd,aps,amsfonts]{revtex4}

\usepackage{graphicx}

\newcommand{\bea}{\begin{eqnarray}}
\newcommand{\ea}{\end{eqnarray}}
\newcommand{\eea}{\end{eqnarray}}

\def\figloc#1#2 {
\begin{figure}\begin{center}
    \includegraphics[width=5in]{fig#1.ps}
    \caption{Figure #1. #2}
    \end{center}\end{figure}
}

\begin{document} 

\title{ Irrotational, two-dimensional Surface waves in fluids}
\author{ William G.~Unruh}
\affiliation{ 
Department of Physics and Astronomy,
University of British Columbia,
Vancouver, B.C., V6T 1Z1 Canada
}
\date{\today}
\begin{abstract} 
The equations for waves on the surface of an irrotational incompressible fluid
are derived in the coordinates of the velocity potential/stream function. The
low frequency shallow water  approximation for these waves is derived for a
varying bottom topography. Most importantly, the conserved norm for the
surface waves is derived, important for quantisation of these waves and their
use in analog models for black holes.
\\
PACS:
47.90.+a,    % Other topics in fluid dynamics
92.60.Dj,    % Gravity waves, tides, and compressional waves
04.80.y.     % Experimental studies of gravity
\end{abstract}    

\maketitle
%%%%%%%%%%%%%%%%%%%%%%%%%%%%%%%%%%%%%%%%%%%%%%%%%%%%%%%%%%%%%%%%%%%%%%%%%%%%%%%
\section{Introduction}\label{intro}
%%%%%%%%%%%%%%%%%%%%%%%%%%%%%%%%%%%%%%%%%%%%%%%%%%%%%%%%%%%%%%%%%%%%%%%%%%%%%%%

One of the most fascinating predictions of Einstein's theory of general 
relativity is the potential existence of black holes -- i.e.\ space-time 
regions from which nothing is able to escape.
Perhaps no less interesting are their antonyms: white holes which 
nothing can penetrate.
Both are described by solutions of the Einstein equations and are related to each 
other via time-inversion, see e.g.\ \cite{misner,ellis}.
 
It is equally fascinating that  some of the predictions for fields in a black
hole spacetime can be modelled by waves in a variety of other situations, with
the interior  of the black hole or white hole horizons can be  mimicked by fluid flow which exceeds the
velocity of the waves in some regions. One of these is the use of surface
waves on a incompressible fluid\cite{schuetz}.  One can alter the flow properties of the
fluid by placing obstacles into the bottom of a flume (a long tank along
which the water flows) to speed up and slow down the fluid over these
obstacles.

One of the difficulties in the theoretical treatment of such systems is the
complicated boundary conditions on the bottom of the tank (where the fluid
velocities must be tangential to the bottom) and the top (where the pressure
of the fluid must be zero or at a constant atmospheric constant pressure). In
fact as we will see the equations for the fluid itself are remarkably simple.
The interesting physics arises entirely from those boundary condition.

We will be interested in irrotational, incompressible flow. While both are
certainly approximations for water flow (the former assumes no turbulence, and
no viscosity which would create vorticity at the shear layer along the bottom,
while the latter assumes that the velocity of sound in the fluid is far higher
than any other velocities in the problem). While this problem has been
investigated before\cite{waves}, this is in general in the three dimensional
context (which is more difficult) and using approximations and expansions for
the shape of the bottom. 

I will assume that the fluid flow is a two dimensional flow-- ie is uniform
across the tank and that the tank maintains a constant width throughout. This
is much simpler case than three dimensional flow, which allows the coordinate
transformations I use. 

 The usual spatial 
coordinates are $x,y$ with $x$ being the horizontal direction in which the
fluid flows, and $y$ is the vertical direction (parallel to the gravitational
acceleration, $g$,
directed in the negative y direction. 

The Euler-Lagrange equations are 
\bea
\partial_t \vec{v} +\vec{v}\cdot\nabla \vec{v} =-g\vec{e}_y -\vec{\nabla} {p\over \rho}
\\
\vec\nabla \cdot\vec v=0
\eea
where the second equation is the incompressibility condition. 
In the usual way, if we assume that the flow is irrotational, then
\bea
\vec v=\vec{\nabla}\tilde\phi
\eea
And the above equation can be written as
\bea
\vec\nabla\left( \partial_t\tilde\phi+{1\over 2} v^2 +gy +{\tilde p\over \rho}\right)=0
\\
\nabla^2\tilde\phi=0
\eea
where $\tilde p$ is the pressure. Let me define the specific pressure, $p={\tilde
p\over \rho}$

In the following I will consider only flows in the $x-y$ directions.
Everything is assumed to be independent of $z$.
Consider the vector $\vec w=\vec{e}_z\times \vec v$. This vector also obeys
\bea
\nabla\cdot \vec w= -\vec{e}_z\cdot \vec\nabla\times\vec v=0\\
\nabla\times\vec w= \vec e_z \vec \nabla \cdot \vec v -
(\vec{e}_z\cdot\vec\nabla)\vec v=0
\eea
since nothing depends on $z$.

Thus we can define 
\bea
\vec w=\vec \nabla \tilde \psi
\eea
where $\tilde \psi$ also obeys $\nabla^2 \tilde\psi=0$
and where 
\bea
\nabla\tilde \psi\cdot\nabla\tilde\phi=0\\
\nabla\tilde\phi\cdot\nabla\tilde\phi=\vec v\cdot\vec v=v^2\\
\nabla\tilde\psi\cdot\nabla\tilde\psi=\vec w\cdot\vec w=v^2\\
\eea

Let me now define a new coordinate system. I could use $\tilde\phi$ and
$\tilde\psi$,
but I will be interested in fluid flows where the velocity approaches a
constant value $v_x=v_0,~~v_y=0$ at large distances. I will thus instead  use the functions
$\psi,\phi$ defined by
\bea
\phi={\tilde\phi\over v_0}\\
\psi={\tilde\psi\over v_0}
\eea
 as
the new coordinates. This choice will also allow me to take the limit as the
velocity $v_0$  goes to zero, where the potentials $\tilde \phi,~\tilde\psi$ are
undefined. Thus at large distances, $\phi=x$ and $\psi=y$.  The spatial metric in the
$xy$ coordinates is 
\bea
ds^2=dx^2+dy^2=g_{ij} dz^i dz^j
\eea
(where the Einstein summation convention has been used where a repeated index
implies summation over that index, and where $z^1=x,~z^2=y$). Do not confuse
$z^i$ with the horizontal direction $z$ which nothing depends on.  The Laplacian is
for a general metric function of $g_{ij}(z)$ is 
\bea
\nabla^2 = {1\over\sqrt{ |g|}}\partial_i \sqrt{|g|}g^{ij} \partial_j
\eea
where $g^{ij}$ are the components of the matrix with is the inverse to the
matrix of coefficients $g_{ij}$ and where $g$ is the determinant of the matrix
with coefficients $g_{ij}$. For a reference regarding metrics and the
coordinate independent equations see almost any book on General
Relativity\cite{wald}.

In two dimensions, if $g_{ij}= f \tilde g_{ij}$ where $f$ is some function of
the coordinates $z^i$, then since
$g^{ij}= {1\over f} \tilde g^{ij}$ and $g=det(g_{ij})= f^2 det(\tilde
g_{ij})=f^2\tilde g$, we have 
$\nabla^2= {1\over f}\tilde\nabla^2$. Metrics such as $g_{ij}$ and
$\tilde g_{ij}$ are said to be confomally related.

Recalling that the change in the metric components from one coordinate system
$z^i$ to a new system $\hat z^j$ are given by 
\bea
\hat g^{kl}= {\partial\hat z^K\over \partial z^i}{\partial z^l\over\partial
z^j} g^{ij}
\eea
where the Einstein summation convention has been used, the components of the
upper components of the usual flat space  metric in this new $\hat z^1=\phi,\hat z^2=\psi$ coordinate
system are
\bea
\hat g^{\phi\phi}=\vec\nabla\phi\cdot\vec\nabla\phi={v^2\over v_0^2}\\
\hat g^{\psi\psi}=\vec\nabla\psi\cdot\vec\nabla\psi={v^2\over v_0^2}\\
\hat g^{\phi\psi}=\vec\nabla\phi\cdot\vec\nabla\psi=0
\eea
Ie, the new  metric (the inverse of this upper form metric) in these new coordinates is a conformally flat metric 
\bea
\hat g_{ij}= {v_0^2\over v^2}\left(\begin{array}{ccc}1&0\\0&1\end{array}\right)
\eea
Since in the $xy$ coordinates the metric is flat, this  metric is also  flat in
$\psi,\phi$ coordinates, (the curvature is not changed by a coordinate
transformation) and the scalar curvature in this new coordinate system is
zero. Using the equation for the scalar  curvature of a metric (and in two
dimensions, the scalar curvature is the only independent component of the
curvature) one gets
\bea
(\partial_\phi^2+\partial_\psi^2 )\ln({v^2\over v_0^2})=0
\eea
(This is valid as long as $v^2\over v_0^2$ is not equal to zero anywhere)

I define 
\bea
\tilde\nabla^2= \partial_\phi^2+\partial_\psi^2
\eea
The Laplacian
\bea
{1\over \sqrt(\hat g)}\partial_i\sqrt{\hat g} \hat g^{ij}\partial_j \Phi
\eea
is, since the metric in $\psi,\phi$ coordinates is  conformally flat, just
\bea
{v^2\over v_0^2} \tilde\nabla^2 \Phi
\eea
for any scalar function $\Phi$.

Since in $x,y$ coordinates, the Laplacian of both the scalar functions
$x$ and $y$ are zero, they must also be zero in $\phi,\psi$ coordinates ( since
the Laplacian is an invariant scalar operator), and, as functions of
$\phi,\psi$, we have
\bea
\tilde\nabla^2 x(\phi,\psi,t)=\tilde\nabla^2 y(\phi,\psi,t)=0
\eea
as the equations of motion obeyed by $x$ and $y$ in these new coordinates.

$\psi$ is the stream function, and the vector $\vec v$ is tangent
to the surfaces of constant $\psi$. $\vec v\cdot \vec\nabla\psi= \vec
v\cdot\vec w=0$.
The bottom of the flow must be tangent to the flow vector (no flow can
penetrate the bottom), and thus must
be a surface of constant $\psi$, which I will take to be $\psi=0$. Similarly, if the flow is stationary, the top of
the water, no matter how convoluted, must also lie along a streamline, since a
particle of the fluid which is at the top, must flow along the top (the
velocity of the particles must be parallel to the top surface). This means that the
top of a stationary flow ( but not a time dependent flow) also is at a
constant value of  $\psi$ which I will label $\psi_T$.

We also have
\bea
\partial_x\phi=\partial_y\psi={v_x\over v_0}\\
\partial_y\phi=-\partial_x\psi={v_y\over v_0}
\partial_\phi x=\partial_\psi y= {v_xv_0\over v^2}
\\
-\partial_\psi x=\partial_\phi y={v_y v_0\over v^2}
\eea
and thus
\bea
{v^2\over v_0^2}&=&{1\over (\partial_\phi y)^2+(\partial_\psi y)^2}\\
&=&{1\over (\partial_\phi x)^2+(\partial_\psi x)^2}
\eea
Solving for $x$ and $y$  as a function of $\psi,\phi$ , which is just
solving the Laplacian in terms of $\psi, \phi$, gives us the
velocity at all points.

The boundary condition along the bottom for these functions must be that the
velocity along the bottom be parallel to the bottom.
If the bottom has the functional form
$y=F(x)$ then $y(\phi,0)=F(x(\phi,0))$. On the top of the flow, we have the
boundary condition that $p=0$. From the Bernoulli equation for a stationary
flow is
\bea
{1\over 2}v^2+gy+p={\rm const}
\eea
which, if the flow has constant velocity $u$  over a constant depth bottom of
height $h$  far
away from the obstacle, gives the equation for the top of the flow
\bea
{1\over 2}v(\phi,\psi_T)^2+gy(\phi,\psi_T)={1\over 2}v_0^2+gh
\eea
Writing this in terms of $\phi,\psi$ we have the upper boundary condition of
\bea
{v_0^2\over (2((\partial_\phi y(\phi,\psi_T))^2+(\partial_\phi
x(\phi,\psi_T))^2))}
+gy(\phi,\psi_T)={1\over 2}v_0^2+gh
\eea
This is a complicated, non-linear, boundary condition. Thus while the
equations of motion of $x,~y$ are simple (Laplacian equals zero), the physics
is all contained in the boudary conditions.

If we are given $y(x)$ as the equation for the bottom, the solution of the
above non-linear boundary value problem is difficult. However if 
, instead of specifying the lower
boundary, one specifies the shape of the  upper boundary $y(\phi,\psi_T)$, one
can use Bernoulli's equation in these new coordinates  to determine the
$\psi$ derivative of $y$ . Since 
\bea
\partial_\psi y= {v_y\over v^2}\\
\partial_\phi y={v_x\over v^2}
\eea
we have
\bea
v^2= {1\over (\partial_\psi y)^2 +(\partial_phi y)^2}
\eea
and Bernoulli's equation is  $v^2+gy=const$ along the top surface of the fluid
where $p=0$. Soving for $\partial_\psi y$ we get
\bea
\partial_\psi y(\phi,\psi_T) = -\sqrt{ -((\partial^\phi y(\phi,\psi_T))^2+
{1\over v_0^2 +g(y(\infty,\psi_T)-y(\phi,\psi_T))}}
\eea

Any function $H(\psi\phi)$ which is a solution of
$\partial_\psi^2 H+\partial_\phi^2 H=0$ can be expanded in exponentials
$e^{ik\phi}$. We see immediately that the dependence of these modes of $\psi$
must be in terms of $e^{\pm k\psi}$  or equivalently in terms of $cosh(k\psi)
$ and $sinh(k\psi)$ for the $\psi$ dependence.  Thus, since
$y$ obeys that equation, we have 
\bea
y(\phi,\psi)=\int  e^{ik\phi}
\left(\alpha_k\cosh(k(\psi-\psi_T))+\beta_k\sinh(k(\psi-\psi_T))\right) 
\eea
with 
\bea
\alpha_k={1\over 2\pi}\int y(\phi,\phi_T) e^{-ik\phi} d\phi\\
\beta_k= {1\over 2\pi} \int {1\over k}\partial_\psi y(\phi,\psi_T)
e^{-ik\phi}d\phi
\eea
Then at the lower boundary, 
\bea
y(\phi,0)= \int\left[ \hat y(k) { \cosh(k\psi_T)} -\hat{\partial_\psi y}(k)
{\sinh(k\psi_T)\over k}\right] e^{ik\phi}dk
\\
x(\phi,0)= \int \int \left[  \hat{\partial  y}(k) { k\cosh(k\psi_T)} +\hat y(k)
{\sinh(k\psi_T)}\right] e^{ik\phi}dkd\phi
\eea
where 
\bea
\hat y(k)= \int y(\phi,\psi_T) e^{-ik\phi} d\phi\\
\hat{\partial y}(k)= \int \partial_\psi y(\phi,\psi_T) e^{-ik\phi} {d\phi}
\eea
This gives the bottom as a parametric set of functions of $\phi$.

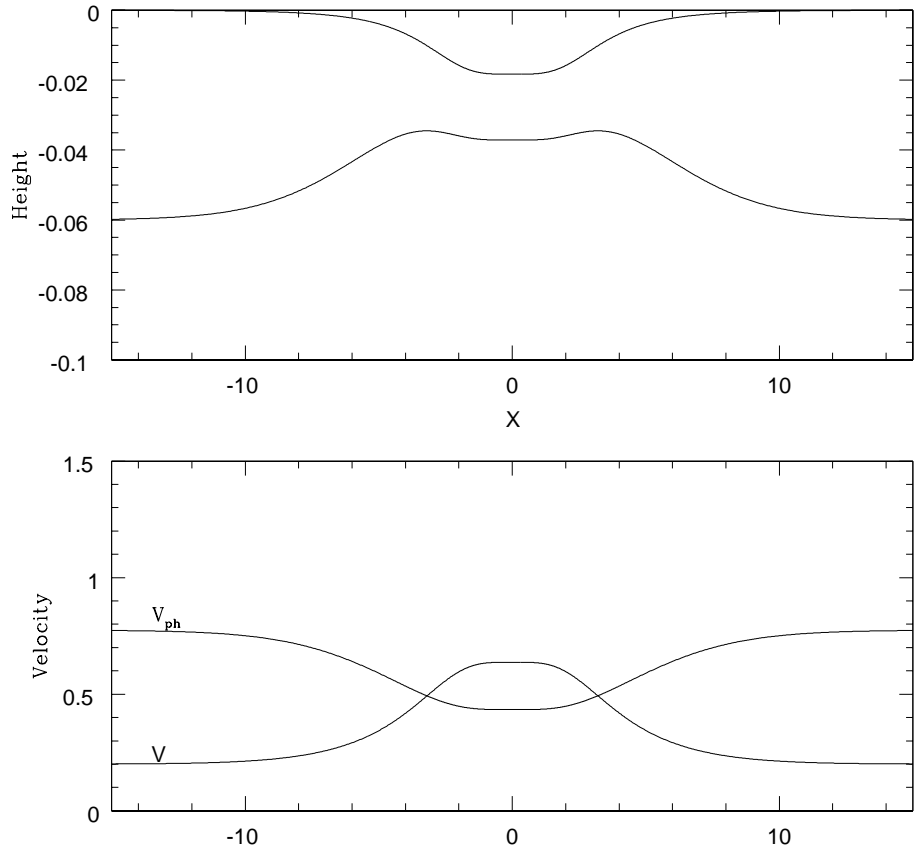
\figloc{1}{ The upper graph gives the top and bottom ($y(\psi_T)$ and $y(0)$
of  a symmetric flume flow with $v_0=.3m/s$. The
top of the flow was specified with $y(\phi,\psi_T)= .015
(e^{(\psi-.5)^2/2}+e^{(\psi+.5)^2/2}$. Note that the bottom of the flume is a
reasonable function of $x$.  The lower graph gives the velocity of the fluid
flow, 
($v(\phi)$) as a
function of $ x$ and the phase velocity of long wavelength waves  $\sqrt{g(y(\phi,\psi_T)-y(\phi,0))}$ as
a function of $x$. The ratio of these two velocities is the Froude number,
which is greater than unity over the obstacle. }

In figure 1 we have an example of sub to supercritical flow over an obstacle. 
calculated as above. Note that the obstacle is a reasonable function  $y(x)$.

\subsection{$v_0=0$ limit}

The boundary condition  equations are easily solved in the limit as 
 $v_0\rightarrow 0$. The upper boundary condition
becomes simply $y=h$ and $\partial_\phi y=0$. This can be solved (in terms of
the unknown lower boundary solutions $y(\phi,0),~x(\phi,0)$ by
\bea
y(\phi,\psi)&=& \int \alpha_k e^{ik\phi}{sinh(k(\psi_T-\psi))\over sinh(k\psi_T)} dk\\
x(\phi,\psi)&=&i\int  \alpha_ke^{ik\phi}{cosh(k(\psi_T-\psi))\over sinh(k\psi_T)} dk
\eea
where
\bea
\alpha_k={1\over 2\pi}\int y(\phi,0)e^{-ik\phi}d\phi
\eea
Of course, we are not given $y(\phi,0)$ but rather $y(\phi,0)=F(x(\phi,0))$.
However one can get rapid convergence by iteration
\bea
x_0(\phi,0)=\phi\\
y_{i+1}(\phi,0)=F(x_i(\phi,0))
\eea
which gives via the above equations the solution $y_{i+1}(\phi,\psi)$ and thus
\bea
x_{i+1}(\phi,0)=\int \partial_\psi y_{i+1}(\phi,0)d\phi 
\eea

For small $v_0$, one can get a first order correction for the surface value
of $y(\phi,\psi_T)$ by taking 
\bea
y(\phi,\psi_T)=h-v_0^2 {1\over (\partial_\psi
y_{v_0=0}(\phi,\psi))^2\vert_{\psi=\psi_T} }
\eea

Ie, for slow flow over a bottom boundary, the stationary solution for that
flow is easy to find. 

\subsection{Formal General solution}

The general solution to the equation $\tilde\nabla^2 F$=0 can be written as
\bea
F=f(\phi+i\psi)+g(\phi-i\psi)
\eea
If $F$ is real, then $g(\phi-i\psi)=(f(\phi+i\psi))^*$
We then have
\bea
x(\phi,\psi)= \hat x(\phi+i\psi)+\hat x^*(\phi+i\psi)\\
y(\phi,\psi)=i(\hat x(\phi+i\psi)-\hat x^*(\phi+i\psi))
\eea
Given the boundary conditions along the bottom, we have
\bea
\hat x(\phi) ={1\over 2}(x_0(\phi,0)-iy_0(\phi,0))
\eea

This of course still leaves the highly non-linear boundary conditions at the
top to solve to find $x$ and $y$ everywhere.

%%%%%%%%%%%%%%%%%%%%%%%%%%%%%%%%%%%%%%%%%%%%%%%%%%%%%%%%%%%%%%%%%%%%%%%%%%%%%%%
\section{Fluctuations}\label{model}
%%%%%%%%%%%%%%%%%%%%%%%%%%%%%%%%%%%%%%%%%%%%%%%%%%%%%%%%%%%%%%%%%%%%%%%%%%%%%%%

Let us assume that we have a background solution to the stationary equation,
$x_0(\phi,\psi), y_0(\phi,\psi)$, or equivalently, $\phi_0(x,y), \psi_0(x,y)$. We want to find the equations for small
perturbations around this background flow.  Let us also consider a solution to
the full time dependent equations,$\phi(x,y,t),\psi(x,y,t)$ together with
their inverses, $x(\phi,\psi,t),y(\phi,\psi,t)$, such that
$y(\phi(x,y,t),\psi(x,y,t),t)=y$ and $x(\phi(x,y,t),\psi(x,y,t),t)=x$. Define the small deviations from the
background by 
\bea
\delta \phi&=&\phi(x,y,t)-\phi_0(x,y)\\
\delta \psi&=&\psi(x,y,t)-\psi_0(x,y)\\
\delta x&=&x(\psi,\phi,t)-x_0(\phi,\psi)\\
\delta y&=&y(\phi,\psi,t)-y_0(\phi,\psi)\\
\eea
Then we have
\bea
y&=&y(\phi_0(x,y)+\delta\phi(x,y,t),\psi_0 (x,y)+\delta\psi(x,y,t)
,t)\\
&=&y_0(\phi_0(x,y)+\delta\phi(x,y,t),\psi_0(x,y)+\delta\psi(x,y,t) ,t) +\delta
y(\phi_0(x,y)+\delta\phi(x,y,t),\psi_0(x,y)+\delta\psi(x,y,t) ,t)
\eea
Keeping terms only to first order in "$\delta$", we have
\bea
y&=&y_0(\phi_0(x,y),\psi_0(x,y)) +\partial_\phi y_0(\phi_0(x,y),\psi_0(x,y)) \delta\phi +\partial_\psi
y_0(\phi_0(x,y),\psi_0(x,y))\delta\psi +\delta y(\phi_0(x,y),\psi_0(x,y))
\eea
{\rm or}
\bea
\delta y(\phi_0(x,y),\psi_0(x,y))&=&-{v_0v_y\over v^2}\delta\phi(x,y) -{v_0v_x\over
v^2} \delta\psi(x,y)
\eea
(where all velocity components are those in the background flow).

Similarly 
\bea
\delta x= -{v_0v_x\over v^2}\delta\phi(x,y) +{v_0v_y\over
v^2} \delta\psi(x,y)
\eea
and 
\bea
\delta\phi(x_0(\phi,\psi),y_0(\phi,\psi))={1\over v_0}( v_x\delta x(\phi,\psi)+v_y \delta
y(\phi,\psi))\\
\delta\psi(x_0(\phi,\psi),y_0(\phi,\psi))= {1\over v_0}(-v_y\delta x(\phi,\psi)+v_x \delta
y(\phi,\psi))
\eea

The Bernoulli equation is 
\bea
v_0\partial_t \phi(x(\phi,\psi,t),y(\phi,\psi,t),t) +{v_0^2\over 2} {1\over
(\partial_\phi x(\phi,\psi,t))^2
+(\partial_\phi y(\phi,\psi,t))^2} +gy(\phi,\psi,t)+p=const
\eea
where the first $\partial_t$ is defined as the derivative keeping $x,y$ fixed,
not $\phi,\psi$ fixed. Here $p$ is the specific pressure.

Writing this equation perturbatively, we have
\bea
-v_x\partial_t\delta x -v_y \partial_t y  - {v_0^2\over( (\partial_\phi
x)^2+(\partial_\phi y)^2)^2 } \left({v_0v_x\over v^2}\partial_\phi \delta x
+{v_0v_y\over
v^2}\partial_\phi y\right) +g\delta y +\delta p=0
\eea
where all of the velocities are the values of the background velocities at the
location $\phi,\psi$. Ie,
$v_x(\phi,\psi)=v_{0x}(x_0(\phi,\psi),y_0(\phi,\psi))$.

We can now rewrite this equation in terms of
$\delta\phi=\delta\phi(x_0(\phi,\psi),y_0(\phi,\psi))$
to get
\bea
v_0\tilde\partial_t\delta\phi +v^2 \left( v_x
\partial_\phi({v_x\over v^2}\delta\phi-{v_y\over
v^2}\delta\psi)+v_y(\partial_\phi({v_y\over v^2}\delta\phi +{v_x\over
v^2}\delta\psi )\right) -g({v_0v_x\over v^2}\delta\psi +{v_0v_y\over v^2}\delta\phi
+\delta p=0
\eea
Recalling that $\partial_\phi {v_x\over v^2}=\partial_\phi\partial_\psi y_0
=\partial_\psi {v_0v_y\over v^2}$ and $\partial_\phi {v_0v_y\over v^2}
=-\partial_\psi {v_0v_x\over v^2}$,
 we finally get

\bea
v_0\tilde \partial_t \delta\phi +v^2\partial_\phi \delta\phi
+\partial_\phi({1\over 2}v^2+gy_0) \delta\phi -\partial_\psi(gy_0+{1\over
2}v^2)\delta\psi +\delta p=0
\eea

The boundary conditions at the bottom are that $\delta x$ and $\delta y$ must
be parallel to the bottom, or $v_x\delta y -v_y\delta x=0$ which is just
\bea
\delta\psi(\phi,0)=0
\eea
At the top, the pressure at the surface must be 0. However the surface is no
longer simply $\psi=\psi_T$ because of the time dependence of the equations. 
Let us assume that the surface is defined by 
\bea
\psi=\Psi(\phi,t)+\psi_T
\eea
Since a particle of the fluid which starts on the surface, remains on the
surface, we can define the fluid coordinates $\eta,\zeta$. Then the velocity
of the fluid is 
\bea
v^\phi= {d\over dt}\phi(\zeta,\eta,t) \\
v^\psi={d\over dt}\psi(\zeta,\eta,t)
\eea
Along the surface, we therefore have
\bea
v^\psi =\partial_t\Psi +v^\phi \partial_\phi\Psi
\eea
But, 
\bea
v^\phi&=&{d\over dt}\phi(x(\eta,\zeta,t),y(\eta,\zeta,t),t)\\
&=& v_x\partial_x \phi +v_y\partial_y \phi +\partial_t \phi\\
&=& {v^2\over v_0}+\partial_t\phi(x,y,t)
\eea
\bea
v^\psi =\partial_t\psi(x,y,t)
\eea
Thus, assuming that $\Psi$ is also small (the same order as the other
"$\delta$" terms), we have
\bea
v_0\partial_t\Psi+{v^2\over v_0}\partial_\phi\Psi=v_0\partial_t\delta\psi
\eea
 
On the surface, we have the Bernoulli equation, which  to first order is
\bea
{1\over 2}v^2(\phi,\psi_T+\Psi)& +& gy_0(\phi,\psi_T+\Psi)-{1\over 2}v^2(\phi,\psi_T+\Psi) +
gy_0(\phi,\psi_t+\Psi)+\tilde\partial_t\delta\phi\\
 &+& v^2\partial_\phi\delta\phi
-\partial_\phi( {1\over 2}v^2+gy)\delta\phi -\partial_\psi({1\over
2}v^2+gy)\delta\psi +p-p_0=0
\eea
But along the surface $\psi=\psi_T$, the background ${1\over 2}v^2+gy$ is constant,
so the $\phi$ derivative is 0. We have
\bea
(\tilde\partial_t+v^2\partial_\phi)\delta\phi +\partial_\psi({1\over
2}v^2+gy)(\Psi-\delta\psi) =0
\eea

Dividing by $G=\partial_\psi({1\over 2}v^2+gy)$ and taking the derivative
$\tilde\partial_t +{v^2\over v_0}\partial_\phi$ we get
\bea
(\tilde\partial_t +{v^2\over v_0}\partial_\phi)\left[{1\over G}(\tilde\partial_t
+{v^2\over v_0}\partial_\phi)\right]\delta\phi- {v^2\over v_0}\partial_\phi\delta\psi=0
\eea
as the equation of motion for the surface wave. $\delta\phi$ and $\delta\psi$
are related by the boundary condition $\delta\phi=0$ along the bottom. 

Since both $\delta\phi$ and $\delta\psi$ obey
$\nabla^2\delta\psi=\nabla^2\delta\phi=0$, we have
\bea
\tilde\nabla^2\delta\psi=\tilde\nabla^2\delta\phi=0
\eea
Furthermore, since
\bea
\partial_x\delta\phi=\partial_y\delta\psi\\
\partial_y\delta\phi=-\partial_x\delta\psi
\eea
so
\bea
\partial_\phi\delta\phi&=&\partial_\phi x_0 \partial_x\delta\phi
+\partial_\phi y_0 \partial_y\delta\phi\phi\\
&=&\partial_\psi y_0\partial_y\delta\psi 
-\partial_\psi x_0 (-\partial_x\delta\psi)=\partial_\psi\delta\psi\\
\partial_\psi\delta\phi&=&-\partial_\phi\delta\psi
\eea

For irrotational time-independent flow, the acceleration of a parcel of
fluid is $\vec v\cdot\nabla \vec v= \vec \nabla ({1\over 2} v^2)$ and the
orthogonal component of this, the centripetal acceleration is 
\bea
{1\over |\nabla \psi|^2} \nabla\psi\cdot\nabla({1\over 2} v^2={1\over
v}\partial_\psi ({1\over 2}v^2)
\eea
Also $g\partial_\psi y= g {v_xv_0\over v^2} \approx g v_0/v$ so$Gv/v_0$ is the effective
gravitational field orthogonal to the flow lines (including the centripital
acceleration) . 

 However it is
important to note that it is the
effective force of gravity only at the surface of the fluid, not at the obstacle to the flow along the
bottom, that is important for the equations of motion. 

\section{Shallow water waves}

Since $\phi,~\psi$ are real functions, the solutions can be written as
\bea
\delta\phi(\phi,\psi)= Z(\phi+i\psi) +(Z(\phi+i\psi))^*\\
\delta\psi(\phi,\psi)=i(Z(\phi+i\psi) -(Z(\phi+i\psi))^*)
\eea
for some function $Z$. These functions clearly satisfy the Laplacian equation
for, and furthermore also satisfy the differential relations on the
derivatives of $x,y$ with respect to $\phi,psi$
This gives
\bea
0=\delta\psi(\phi,0)= i(Z(\phi)-Z^*(\phi))
\eea
Ie, $Z$ is a real function of a  real arguments.
which gives
\bea
\delta\phi(\phi,\psi)=(Z(\phi+i\psi)+Z(\phi-i\psi))\approx 2 Z(\phi)+Z''(\phi)\psi^2
\\
\delta\psi= 2\psi Z'(\phi)
\eea
or, to first order in $\psi_T$
\bea
\delta\psi=\psi_T \partial_\phi \delta\phi
\eea
The equation for the waves then becomes
\bea
(\tilde\partial_t +v^2\partial_\phi){1\over G} (\tilde\partial_t
+v^2\partial_\phi)\delta\phi -v^2 \psi_T \partial_\phi^2\delta\phi=0
\eea

We note that this is not a Hermitian operator acting on $\delta\phi$. Recall
that a Hermitian operator is one such that 
\bea
\int \delta \hat \phi {\cal H} \phi d\phi dt = \int ({\cal H} \delta\hat\phi)
\delta \phi d\phi dt
\eea
if we assume  that all of the boundary terms in the integration by parts are
zero.  We can rewrite the equation for $\delta\phi$  by dividing by $v^2$ as  
\bea
(\tilde\partial_t +\partial_\phi v^2){1\over v^2G} (\tilde\partial_t
+v^2\partial_\phi)\delta\phi - \psi_T \partial_\phi^2\delta\phi=0
\eea
This is a symmetric equation, derivable from an action,
\bea
\int \left[ {1\over v^2G}(\tilde\partial_t
+v^2\partial_\phi)\delta\phi^* (\tilde\partial_t
+v^2\partial_\phi)\delta\phi
-\Psi_T\partial_\phi\delta\phi^*\partial_\phi\delta\phi\right] d\phi dt 
\eea
 This action has the global symmetry $\delta\phi\rightarrow
e^{i\mu}\delta\phi$ and thus  has the usual Noether current associated with
this symmetry. In particular it has the  conserved norm

\bea
<\delta\phi,\delta\phi'>={i\over 2}\int\left\{ \delta\phi^* {1\over
Gv}(\partial_t+v^2\partial_\phi){\delta\phi'\over v}- \delta\phi'
{1\over
Gv}(\partial_t+v^2\partial_\phi){\delta\phi^*\over v}\right\} d\phi
\eea

\section{ Deep Water waves}
For deep water waves, we can assume that either
$Z(\phi+i\psi_T)>>Z(\psi-i\phi_T)$
or $Z(\phi+i\psi_T)<<Z(\psi-i\phi_T)$. (ie, we assume that as analytic functions,
$Z$ goes to zero either in the upper or lower half plane.)

Let us also assume  it is the first case, and let us define $\hat
Z(\phi)=Z(\phi+i\psi_T)$, and that $\tilde\partial_t\delta\phi=i\omega\delta\phi$ We then
have
\bea
(i\omega +v^2\partial_\phi){1\over G} (i\omega+v^2_\phi)]\hat Z
-(-i)v^2\partial_\phi \hat Z =0
\eea
If we assume that $K=i(\partial_\phi ln(\hat Z))$ is large and negative, such
that $\hat Z $ varies faster than $v^2$ or $G$, we have approximately
\bea
{(\omega+v^2(\phi)K)^2\over G}+Kv^2=0
\eea
or
\bea
\omega=-v^2K \pm \sqrt{v^2GK}
\eea

\section{General Linearized waves}

The equation in general is
\bea
(\tilde\partial_t+v^2\partial_\phi){1\over G}
(\tilde\partial_t+v^2\partial_\phi)\delta\phi - v^2\partial_\phi \delta\psi =0
\eea
Fourier transforming with respect to $\phi$ and $\psi$, and using the fact that $\delta\psi=0$
at $\psi=0$, the functions $\delta\phi,\delta\psi$   then can be written as
\bea
\delta\phi(\phi,\psi,t)=\int A(k,t)e^{ik\phi}\cosh(k\psi)dk\\
\delta\psi(\phi,\psi,t)=i\int A(k,t) e^{ik\phi}\sinh(k\psi)dk
\eea
since again they obey the Laplacian equal to zero in these variables.

Defining $B(k,t)=A(k,t)cosh(k\psi)$ this can be written as 
\bea
\delta\phi(\phi,\psi_T,t)&=&\int B(k,t) e^{ik\phi} dk\\
\delta\psi(\phi,\psi_T,t)&=&i\int B(k,t) e^{ik\phi_T} \tanh(k\psi) dk =i
\tanh(-i\psi_T\partial_\phi) \int
B(k,t) e^{ik\phi}dk=i \tanh(-i\psi_T\partial_\phi)\delta\phi
\eea
Thus the equation of the surface waves can be written as
\bea
0&=& (\tilde\partial_t+\partial_\phi v^2){1\over v^2 G}
(\tilde\partial_t+v^2\partial_\phi)\delta\phi
-i\partial_k\tanh(-i\psi_T\partial_\phi) \delta\phi
\eea

Ie, we get the usual  $\tanh$ dispersion relation for the transition from shallow to
deep water waves. 

This equation is symmetric and real, and thus if $\delta\phi$ is a solution,
so is $\delta\phi^*$. Again this gives a conserved norm between two
solutions to the equations of motion $\delta\phi$ and $\delta\phi'$ of
\bea
<\delta\phi,\delta\phi'>=\int {1\over v^2 G}\left[\phi^*
(\tilde\partial_t+v^2\partial_\phi)\delta\phi' 
-\delta\phi'(\tilde\partial_t+v^2\partial_\phi) {\delta\phi}^*\right]d\phi
\eea

We note that this equation depends only the conditions at the surface of the
flow. It is defined entirely in terms of the factors ($v^2$ and
$G=\partial_\psi (gy+{1\over 2}v^2$) defined at $\psi=\psi_T$, and is
independent of the obstacles, or the flow throughout the rest of the stream
except insofar as they affect the flow at the surface. 
This might well change if either vorticity or viscosity were introduced into
the equations. 

This norm is crucial to the analysis of the wave equation.  It is conserved (in
the absence of viscosity), and in the use of such waves as models for black
holes, it is this norm which determines the Bugoliubov coefficients (or the
amplification factor) for waves in the vicinity of a horizon (blocking flow in
the hydrodynamics sense) and determines the quantum noise (Hawking radiation)
emitted by such a horizon analog.  The quantum norm used in the quantization procedure
is 
\bea
\langle \delta\phi,\delta\phi\rangle_Q= {i\over 2}<\delta\phi,\delta\phi>
\eea

If we define a new coordinate $\hat\phi=\int {1\over v^2} d\phi$, the norm
becomes
\bea
<\delta\phi,\delta\phi'>=  \int({1\over G}\left[\delta\phi^*(\tilde\partial_t -
\partial_{\hat\psi})\delta\phi' - \delta\phi^*(\tilde\partial_t -
\partial_{\hat\psi})\delta\phi \right]d\hat\phi
\eea

If the surface of the flow is shallow (${dy_T\over dx}<<1$) then ${d\phi\over
dx} =v_x\approx v$ and $\hat\phi\approx {dx\over v_x}$.  

To relate this to the measured quantity, the vertical displacement at the
surface of the waves, we must relate $\delta\phi$ to $\delta y$ at the surface
of the fluid. We have
\bea
\Psi(t,\phi)= \psi(t,x,y_T(t,x)) -\psi_T
=\delta\psi(t,x(\phi,\psi_T),y(\phi,\psi_T) +v_x \delta y_T
\eea
or
\bea
\delta y_T= {1\over v_x}(\Psi-\delta\psi) ={1\over G v_x} \partial_t
+v^2\partial_\phi)\delta\phi
\eea
Now, $G v_x\approx g {v_x^2\over v^2}\approx g$ (ignoring the centrifugal
contribution to the effective gravity),  so the norm becomes
\bea
<\delta y_t,\delta y_T>&=& \int {v^2\over  g} \left[(
\partial_t+\partial_{\hat\psi})^{-1} \delta
y_T^* \delta y_T - ( \partial_t+\partial_{\hat\psi})^{-1} \delta
y_T  \delta y_T^*\right]d\hat\phi\\
&=& \int {1\over  g} \left[(( \partial_t+\partial_{\hat\psi})^{-1} \sqrt{v}\delta
y_T^*) \sqrt{v}\delta y_T - (( \partial_t+\partial_{\hat\psi})^{-1} \sqrt{v}\delta
y_T) \sqrt{v} \delta y_T^*\right]d\hat\phi
\eea
and $d{\hat\phi}{d\phi\over v^2}\approx {dx\over v}$

If we assume that the incoming wave is at a set frequency $\omega$ and take the fourier transform with respect to $t,\hat x$ of
$\sqrt{v(\hat\psi)}y_T(t,\hat\phi)$ this becomes
\bea
<\delta y,\delta y>=\int {|(\sqrt{v} y_T)(\hat k)|^2\over (\omega +\hat k)} d\hat k
\eea

We can also look at the norm current. 
\bea
&&\partial_t \int_{\phi_1}^{\phi_2} {1\over v^2 G}\left[\phi^*
(\tilde\partial_t+v^2\partial_\phi)\delta\phi'
-\delta\phi'(\tilde\partial_t+v^2\partial_\phi) {\delta\phi}^*\right]d\phi
\\
&&~= \int_{\phi_1}^{\phi_2} \partial_x({1\over
G}(\tilde\partial_t+v^2\partial_\phi^*)\delta\phi -\partial_x({1\over
G}(\tilde\partial_t+v^2\partial_\phi)\delta\phi^*)
+\left[(-i\partial_\phi\tanh(-i\psi_T
\partial_\phi)\delta\phi^*)\delta\phi-(i\partial_\phi\tanh(i\psi_T)
\partial_\phi)\delta\phi^*)\delta\phi \right] d\phi
\nonumber
\eea
The integrand is a  complete derivatives.  Although this is not obvious for
the terms  with the $\tanh$ in
them, we can use
 \bea
(\partial_\phi^{2n}\delta\phi^*)\delta\phi - \delta\phi^*
\partial_\phi^{2n}\delta\phi =\partial_\phi (\sum_{r=0}^{2n-1}(-1)^{r}
\partial_\phi^r \delta\phi^{*r} \partial_\phi^{2n-1-r}\delta\phi)
\eea
and the fact that 
 $i\partial_\phi\tanh(i\psi_T \partial_\phi)$ can be expanded in a power
series in 
$\partial_\phi^2$ to show that they also a complete derivative..

Thus the integrand can be written in terms of a complete derivative of with
respect to $\partial_\psi$ and  we can regard the term that is being taken
the derivative of as a spatial norm current $J^\phi$ so that if $J^t$ is the
temporal part of the norm current, we have $\partial_t J^t+\partial_\phi
J^\phi=0$.

If we are in a regime where $\delta\phi =A e^{-i\omega t -k\phi}$, (ie, a regime
where the velocity $v$ and $G$ are both constants), then we have
\bea
J^\phi=  i|A|^2 {(\omega + v^2k)\over Gv^2}+ \partial_k(k\tanh(\Psi_T k))
=i|A|^2 \omega{(1
+v^2/v_p -2v_g)\over Gv^2}
\eea
where $v_p$ and $v_g$ are the phase and group velocity of the wave.  In a
situation in which one has a wave train with some definite frequency and wave
number  entering
a region, then the sum of all the norm currents for each $k$
at the boundary of the region must be zero.

\section{Blocking flow}
Let us return to the static situation. Define $U=\partial_\phi \delta\phi$, we
have the equation
\bea
\partial_\phi {v^2\over G }U +i\tanh(i\psi_T\partial_\phi)U=0 
\eea
As above, there is a solution if we assume that the derivatives are small,
which gives
\bea
U={{\rm const}\over {v^2\over G}-\psi_T}
\eea
For rapid variations, we have
\bea
U={\rm const} {v^2\over G}e^{i\int{G\over v^2}d\phi}
\eea
with the transition from one to the other occuring roughly when the
logarithmic derivatives of the two solutions are equal
\bea
{(v^2/G)'\over {v^2\over G}-\psi_T}\approx {\sqrt {{(v^2/ G)'}^2+1}\over v^2/G}\\
\eea

Defining the Froude number by $F^2={ v^2\over G\psi_T}$ (the square of the
velocity of the fluid over the velocity of the long wavelengths in the fluid
in the WKB approximation), we have
\bea
{(F^2)' \over F^2-1} \approx \sqrt {4 (\ln(F)')^2+ \left({1\over
F^2\psi_T}\right)^2}
\eea

Note that for a non-trivial rate of change of of the bottom, the turning point
occurs well before the horizon. 

The ${}'$ denotes derivative with respect to $\phi$ not x. We can rewrite this
approximately (assuming that ${v_x\over v}\approx 1$ and that $2ln(F)'<1\over
F^2\psi_T$ and $\psi_T\approx v d$ where d is the depth of the water at
postion $x$. 
as 
\bea
{dF^2\over dx} \approx{ (F^2-1) \over F^2 d}
\eea

Note that this transition occurs before $ G\psi_T=v^2$ or Froude number equals
1. The wave on the slope piles up and its frequency makes the transition to
 deep water wave before we hit the effective horizon. 

The long wavelength equation, 
\bea
{1\over v^2} (\tilde\partial_t+v^2\partial_\phi){1\over G}
(\tilde\partial_t+v^2\partial_\phi)\delta\phi
-\psi_T\partial_\phi^2 \delta\phi=0
\eea
is not that of a two dimension metric, which is always conformally flat,  but can be written as a the wave equation for a
three dimensional metric where all derivatives are equal to zero in the third
$\xi$ dimension for the variable $\delta\phi$.
The metric is 
\bea
ds^2=\alpha((1-{v^2\over G\psi_T})dt^2+2{1\over G\psi_T} dtd\phi-{1\over
v^2G\psi_T} d\phi^2) -{1\over v^2 G\psi_T} d\xi^2
\eea
where $\alpha$ is an arbitrary function of $\phi$, a two dimensional conformal
factor which does not affect the two dimensional wave equation
This metric has  surface gravity 
\bea
\kappa= {v^2\over 2}\partial_\phi\left( ({v^2\over G\psi_T})\right)  
={1\over 2}v^2\partial_\phi F^2
\eea
(The surface gravity is the acceleration in the metric  as seen from far away.
for a static time independent metric in a coordinate system which is regular
across the horizon, it can be defined by $\kappa=\Gamma^t_{tt}$ at the
horizon,  where $\Gamma^i_{jk}$ is the
Christofell symbol for the metric. Then $\Gamma^t_{tt}=- {1\over
2}g^{t\phi}(\partial_\phi g_{tt})$ at the horizon.)

\section{Conversion to $\delta y$}

Of course $\delta\phi$ is not what is actually measured in an experiment. That
is the fluctuation $\Delta y(x)$ which is the difference in height between the
stationary flow, and the height with the wave present. 
We can relate this to $\delta\psi$ and $\Psi$.

\bea
y_s(x,t)=y_0(x)+\Delta y(x,t)
\eea
where $y_0$ is the surface for the background.
\bea
\delta y = {v_y\over v^2}\delta\phi+{v_x\over v^2}\delta\psi
\eea
Since $\delta\psi= \tanh(\Psi_H \partial_\phi)\delta\phi$, we have
\bea
v^2\delta y=[v_y+v_x\tanh(\Psi_H \partial_\phi))]\delta\phi
\eea
Inverting this for deep water waves, 
\bea
\delta\phi= {v^2\over v_y +v_x} \delta y
\eea
while for shallow water waves
\bea
\delta\phi= \int exp^{-\int {v_y\over v_x}d\phi}{v^2\over v_x} \delta y d\phi
\eea
The integrand in the exponent is non-zero only in the region where the
background flow is dimpled, and, since $v_y\over v_x$ is in general very
small, the exponential can be neglected in most situations. 

In the intermediate region, where the wave changes from shallow to deep water
wave, there is no easy solution to these equations, but they can be integrated
numerically. 

\section{Waves in stationary water over uneven bottom}

In the limit as $v_0$ goes to zero, so does $v$ with the ratio being a finite
function. 
$y$ obeys the equation $\partial_\phi^2+\partial_\psi^2y=0$ with the boundary
conditions along the bottom that $y=Y(x)$, with Y the given function of x of
the bottom, and along the top, $y=H$, a constant. If we assume that we know 
$Y(\phi)$ (instead of $Y(x)$) along the bottom, this can be solved by 
\bea
y(\phi,\psi)= H+\int \alpha(k) e^{ik\phi} {sinh(k(\psi-\psi_T))\over
sinh(k\psi_T)} dk
\eea
where 
\bea
\alpha(k)= {1\over 2\pi} \int Y(\phi)e^{ik\phi}
\eea
and
\bea
x(\phi,\psi)= \phi +i\int\alpha(k) {cosh(k(\psi-\psi_T))\over
sinh(k(\psi_T))}dk
\eea
One gets rapid convergence if one starts by taking $x=\phi$, substituting into
$Y(x(\phi))$ to find $Y(\phi)$, finding the new $x(\phi)$ and substituting in
again. 

Then $v_y\over v_0$ at the surface is zero, while 
\bea
{v_0\over v_x}={v_0\over
v}=\partial_\psi y=\int k\alpha(k) {1\over
sinh(k(\psi_T))}dk 
\eea

The equation for small perturbations becomes
\bea
{v_0^2\over v^2 G}\partial_t^2 \delta\phi - i\partial_\phi
\tanh(i\psi_T\partial_\phi)\delta\phi=0
\eea
where 
\bea
{v^2\over v_0^2}G={v^2\over v_0^2}g\partial_\psi y=g{v_x\over
v_0}=g{\partial\phi\over\partial_x}
\eea
.If the depth is constant, the backgound $\psi=y$
and $\phi=x$ giving the usual equation, which allows us to write
\bea
\partial_t^2\delta\phi +ig\partial_x \tanh(\psi_T\partial_\phi)\delta\phi
\eea

For deep water waves, where the $\tanh$ is unity, this equation is exactly the
same as the deep water equation for constant depth. The fact that the bottom
varies makes no difference to the propagation of the waves, as one would
expect. 

For shallow water waves, where the $\tanh$ can be approximated as the linear
function in its argument, the equation becomes
\bea
\partial_t^2\delta\phi = \psi_T\partial_x\partial_\phi\delta\psi
=g\psi_T{v_0\over v} \partial_x^2 \delta\phi
\eea
 
This allows us to determine the wave propagation over an arbitrarily defined
bottom. Note that in the stationary limit, the background flow is certainly
irrotational, implying that the assumptions made here should certainly be
valid (of course neglecting the viscosity of the fluid).

%%%%%%%%%%%%%%%%%%%%%%%%%%%%%%%%%%%%%%%%%%%%%%%%%%%%%%%%%%%%%%%%%%%%%%%%%%%%%%%

%%%%%%%%%%%%%%%%%%%%%%%%%%%%%%%%%%%%%%%%%%%%%%%%%%%%%%%%%%%%%%%%%%%%%%%%%%%%%%%

{\bf Acknowledgement}
%\\
This work was supported by The 
Canadian Institute for Advanced Research (CIfAR) and by
The Natureal Science and Engineering Research Council of Canada (NSERC) and
was completed while the author was a  visitor at the Perimeter Institute. 
%
%\addcontentsline{toc}{section}{References}%%%%%%%%%%%%%%%%%%%%%%%%%%%%%%%%
%

%
\end{document}